\documentclass[prl,twocolumn,superscriptaddress,floattfix,aps,final]{revtex4}
\usepackage{graphicx} 
\usepackage{color} 
\usepackage{amsmath} 
\usepackage[urlcolor=blue, hyperindex, colorlinks, bookmarks=true,linkcolor=black,citecolor=black]{hyperref} 
\usepackage{amssymb}

\newcommand{\ket}[1]{\left\lvert #1 \right\rangle}
\DeclareMathOperator{\tr}{Tr}

\newcommand{\MHz}{\mathrm{MHz}}

\newcommand{\us}{\mu\mathrm{s}}
\newcommand{\ns}{\mathrm{ns}}

\newcommand{\mV}{\mathrm{mV}}

\newcommand{\VH}{V_{\mathrm{H}}}

\newcommand{\wL}{\omega^{\mathrm{L}}}
\newcommand{\wR}{\omega^{\mathrm{R}}}

\newcommand{\wcav}{\omega_{\mathrm{C}}}

\newcommand{\chiL}{\chi^{\mathrm{L}}}
\newcommand{\chiR}{\chi^{\mathrm{R}}}

\newcommand{\XL}{X^{\mathrm{L}}}
\newcommand{\XR}{X^{\mathrm{R}}}
\newcommand{\YL}{Y^{\mathrm{L}}}
\newcommand{\YR}{Y^{\mathrm{R}}}
\newcommand{\ZL}{Z^{\mathrm{L}}}
\newcommand{\ZR}{Z^{\mathrm{R}}}

\newcommand{\conc}{\mathcal{C}}
\newcommand{\F}{\mathcal{F}}
\newcommand{\Pauli}{\vec{P}}
\newcommand{\B}{\mathcal{B}}
\newcommand{\W}{\mathcal{W}}

\newcommand{\mXX}{\langle XX \rangle}

\newcommand{\mXZ}{\langle XZ \rangle}

\newcommand{\mZX}{\langle ZX \rangle}

\newcommand{\mZZ}{\langle ZZ \rangle}

\newcommand{\mCHSHzxzx}{\langle \mathbb{C}_{ZXZX} \rangle}
\newcommand{\mCHSHzxxz}{\langle \mathbb{C}_{ZXXZ} \rangle}

\newcommand{\CHSHzzxx}{\mathbb{C}_{ZZXX}}
\newcommand{\CHSHxxzz}{\mathbb{C}_{XXZZ}}
\newcommand{\CHSHzxzx}{\mathbb{C}_{ZXZX}}
\newcommand{\CHSHzxxz}{\mathbb{C}_{ZXXZ}}

\newcommand{\bra}[1]{\langle{#1}|}

\newcommand{\CHSH}{\mathbb{C}}
\newcommand{\mCHSH}{\langle \CHSH \rangle}
\newcommand{\amCHSH}{|\mCHSH|}

\begin{document}

\title{Entanglement Metrology Using a Joint Readout of Superconducting Qubits}
\date{\today}
\author{J. M. Chow}
\affiliation{Departments of Physics and Applied Physics, Yale University, New Haven, Connecticut 06520, USA}
\author{L. DiCarlo}
\affiliation{Departments of Physics and Applied Physics, Yale University, New Haven, Connecticut 06520, USA}
\author{J. M. Gambetta}
\affiliation{Institute for Quantum Computing and Department of Physics and Astronomy, University of Waterloo, Waterloo, Ontario, Canada N2L 3G1}
\author{A. Nunnenkamp}
\affiliation{Departments of Physics and Applied Physics, Yale University, New Haven, Connecticut 06520, USA}
\author{Lev S. Bishop}
\affiliation{Departments of Physics and Applied Physics, Yale University, New Haven, Connecticut 06520, USA}
\author{L. Frunzio}
\affiliation{Departments of Physics and Applied Physics, Yale University, New Haven, Connecticut 06520, USA}
\author{M. H. Devoret}
\affiliation{Departments of Physics and Applied Physics, Yale University, New Haven, Connecticut 06520, USA}
\author{S. M. Girvin}
\affiliation{Departments of Physics and Applied Physics, Yale University, New Haven, Connecticut 06520, USA}
\author{R. J. Schoelkopf}
\affiliation{Departments of Physics and Applied Physics, Yale University, New Haven, Connecticut 06520, USA}

\begin{abstract}
Accurate and precise detection of multi-qubit entanglement is key for the experimental development of quantum computation. Traditionally, non-classical correlations between entangled qubits are measured by counting coincidences between single-shot readouts of individual qubits. We report entanglement metrology using a single detection channel with direct access to ensemble-averaged correlations between two superconducting qubits. Following validation and calibration of this joint readout, we demonstrate full quantum tomography on both separable and highly-entangled two-qubit states produced on demand. Using a subset of the measurements required for full tomography, we perform entanglement metrology with $\sim$$95\%$ accuracy and $\sim$$98\%$ precision despite $\sim$$10\%$ fidelity of single measurements. For the highly entangled states, measured Clauser-Horne-Shimony-Holt operators reach a maximum value of $2.61\pm 0.04$, and entanglement witnesses give a lower bound of $\sim$$88\%$ on concurrence. In its present form, this detector will be able to resolve future improvements in the production of two-qubit entanglement and is immediately extendable to 3 or 4 qubits.
\end{abstract}

\maketitle

Since 1964, when Bell made testable \cite{bell_einstein-podolsky-rosen_1964} the famous paradox formulated by Einstein, Podolsky and Rosen \cite{EPR} questioning non-locality in quantum mechanics, the measurement of correlations between quantum systems has been central to foundational tests against alternate theories. The once `spooky' non-classical correlations known as quantum entanglement have since been amply examined in experiments \cite{CH_bell,aspect_bells_1982,tittel_bells_1998,zeilinger_experiment_1999,ansmann_thesis}. Recently, entanglement has gained prominence as a key resource for large-scale quantum computation, making today's quantum engineer less concerned with foundational issues, and more with the generation and detection of near-perfect entanglement between qubits.

All approaches to metrology of entanglement, such as quantum state tomography, entanglement witnesses and generalized Bell inequality violations \cite{horodecki_rmp}, require experimental measurement of ensemble-averaged qubit-qubit correlations. The paradigm for such detection used in trapped-ion systems~\cite{Haffner_multiparticle_ions_2005,Leibfried:2005yq}, where single-shot qubit readouts with fidelities exceeding $99.99\%$ have been realized \cite{myerson:200502}, is to calculate correlations by measuring coincidences between individual detector `clicks' over many repetitions. In solid-state systems, however, measurement cross-talk and lower readout fidelities make the accurate and precise measurement of  correlations by this approach technically challenging, despite significant improvements in recent years~\cite{martinis_rabi_2002,siddiqi_jba_2004,Lupascu:2007nx}. Various approaches seeking direct access to  correlations with joint, or quadratic,  on-chip detectors have been theoretically proposed for mesoscopic systems  \cite{trauzettel:235331,Mao_mesoQM_2004}.

In this Report, we demonstrate entanglement metrology with a single measurement channel that performs a joint readout of two charge-based superconducting qubits. This readout is based on qubit-state dependent shifts of a microwave transmission-line resonator in a circuit quantum electrodynamics architecture \cite{wallra_strong_2004}. By  validating and calibrating the measurement model for this single-channel detector, we circumvent a low $\sim$$10\%$ single-shot detection fidelity and obtain ensemble-averaged correlations with $\sim$$95\%$ accuracy and $\sim$$98\%$ precision. While this joint readout has been previously used in superconducting qubit systems \cite{filipp_joint_2009}, we demonstrate for the first time its use to detect state-of-the-art highly-entangled states produced on demand, with fidelity to targeted Bell states $>$$90\%$, a lower bound of $\sim$$88\%$ on concurrence, and violation of Clauser-Horne-Shimony-Holt (CHSH) inequalities \cite{clauser_proposed_1969} with a value of $2.61\pm 0.04$. Furthermore, we anticipate that future improvements in entanglement production with this architecture will be detectable with the joint readout in its present form.

\begin{figure*}[t!]
\centering
\includegraphics[scale=1.0]{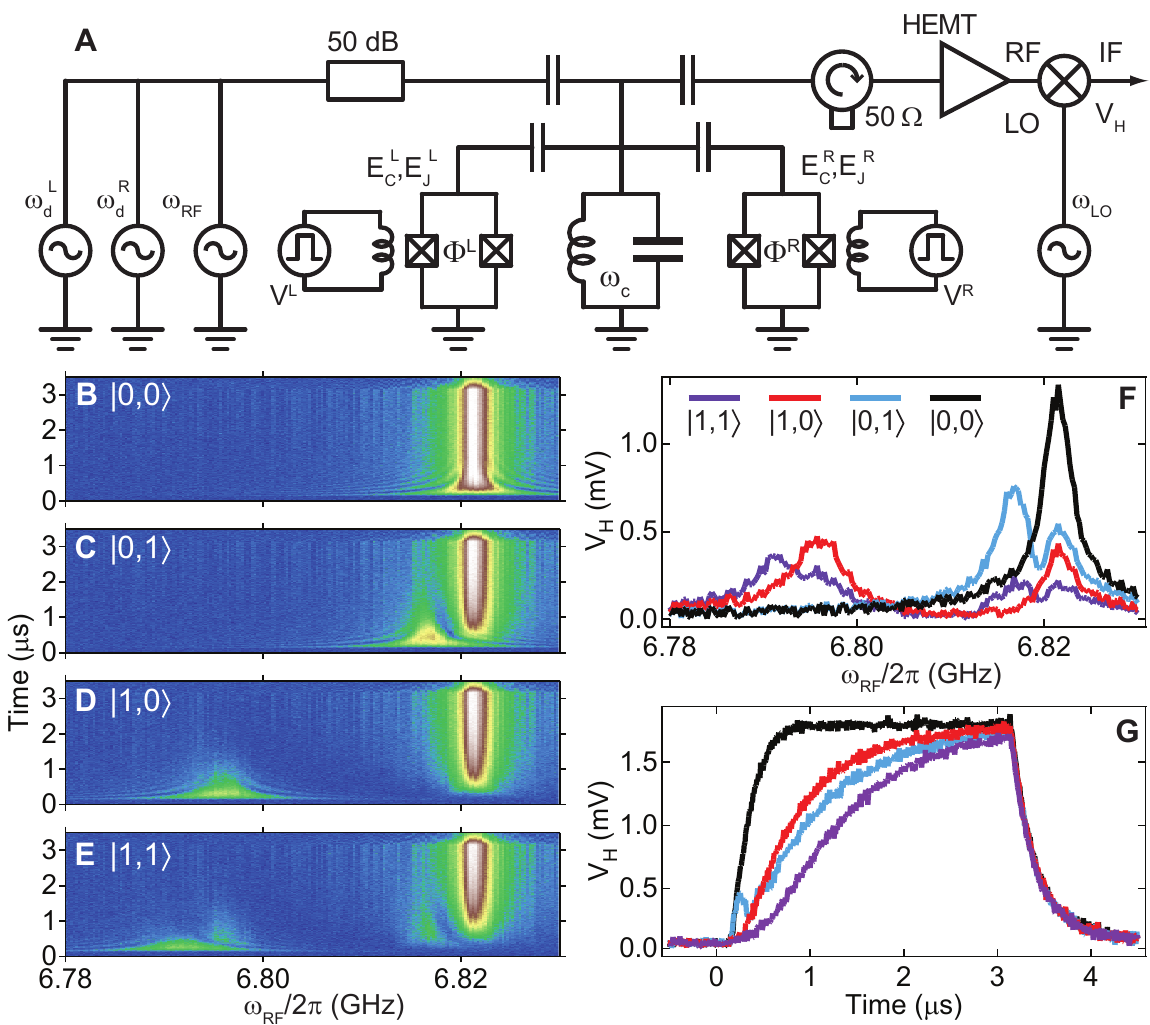}
\caption{Experimental setup and readout. \textbf{(A)} Circuit diagram of experimental setup, representing the cavity as a lumped $L-C$ resonator. Microwave drive tones that address the cavity, $\omega_{\text{RF}}$, and qubits, $\omega_\text{d}^{\text{L(R)}}$, are applied via the cavity input line. The left (right) qubit has charging energy $E_C^{\text{L(R)}}$ and Josephson energy $E_J^{\text{L(R)}}$ (flux $\Phi^{\text{L(R)}}$ tunable via $V^{\text{L(R)}})$. The multiplexed qubit-state information is transmitted out on a single readout line, amplified through a high electron mobility transistor (HEMT) amplifier at $4\, \text{K}$ and mixed down at room temperature for digital processing. \textbf{(B-E)} Color images of transmitted $\VH$ amplitude as a function of time and cavity drive frequency $\omega_{\text{RF}}$, for two-qubit states \textbf{(B)} $\ket{0,0}$, \textbf{(C)} $\ket{0,1}$, \textbf{(D)} $\ket{1,0}$, and \textbf{(E)} $\ket{1,1}$. In this color scheme blue (white) represents $\VH=0\,(1.9)\, \mV$. On short time scales $t<T_1^{\text{L(R)}}$, maximal transmission occurs at different frequencies for the four cases (see text for details). \textbf{(F)} Temporal average of the homodyne voltage transients in \textbf{(B-E)} over the first $500\,\ns$ shows well resolved peaks. \textbf{(G)} Measured $\VH$ transients for drive at the cavity frequency corresponding to $\ket{0,0}$, showing the measurement transient for the four computational states.}
\end{figure*}

Our device consists of a superconducting transmission-line resonator that couples two transmon qubits \cite{koch_charge-insensitive_2007, schreier_suppressing_2008}. Beyond mediating the interaction between the qubits, the bus serves as a single detection channel for their joint readout. Arbitrary single-qubit $x$- and $y$-rotations are performed using in-phase and quadrature microwave pulses resonant with the ground to first-excited state transition of each transmon \cite{chow_bm_2009}. Reduction of leakage to the second-excited state is accomplished through the technique of derivative removal by adiabatic gate \cite{motzoi_2009}, resulting in error rates of $\sim$$1\%$ in single-qubit rotations (See attached supporting material). Local flux-bias lines tune the qubit transition frequencies on nanosecond time scales, allowing control of single-qubit dynamical phases and of a  $Z^{\text{L}} \otimes Z^{\text{R}}$ interaction, both crucial for implementing a two-qubit conditional-phase (C-phase) gate~\cite{dicarlo_2009} (here $Z^{\text{L(R)}}=\sigma_z^{\text{L(R)}}$ is the single-qubit Pauli $z$ operator  \cite{nielsen_chuang_2000} on the left (right) qubit). A schematic of the experimental setup \cite{chow_supp} is shown in Fig.\,1A.

Combining the single-qubit rotations with the C-phase gate, it is possible to produce maximally-entangled states such as the four standard Bell states, $\ket{\Psi_{\pm}} = (\ket{0,0} \pm \ket{1,1})/\sqrt{2}$ and $\ket{\Phi_{\pm}} = (\ket{1,0} \pm \ket{0,1})/\sqrt{2}$, where $\ket{l,r}$ denotes excitation level $l(r)$ of the left (right) qubit \cite{dicarlo_2009}. Yet, to accurately and precisely detect arbitrary two-qubit states, a complete physical model and calibration of the joint readout is necessary.

The physical mechanism enabling the joint readout is a qubit state-dependent dispersive cavity shift that is large relative to the cavity linewidth $\kappa$. In this `strong dispersive' regime \cite{schuster_resolving_2007}, the system is described by a dispersive Tavis-Cummings Hamiltonian
\begin{eqnarray}
	H_{\text{TC}}/\hbar = (\wcav+\chiL Z^{\text{L}} +\chiR Z^{\text{R}})a^{\dag}a-\frac{\wL}{2}Z^{\text{L}} - \frac{\wR}{2}Z^{\text{R}},
\end{eqnarray}
where $\wcav$ is the bare resonator frequency, $\omega^{\text{L(R)}}$ is the Lamb-shifted ground to first excited state transition frequency for the left (right) qubit, and  $\chi^{\text{L(R)}}$ is the left (right) qubit-state dependent cavity shift. Actual parameter values were determined by spectroscopy experiments \cite{chow_supp}. From Eq.\,(1), there can be at most four distinct cavity frequencies corresponding to the joint state of the two qubits. 

The linear dispersive shifts are calibrated by measurement of the transmitted homodyne voltage $\VH$ transient when pulsing a measurement tone, having prepared each of the four computational basis states, $\ket{0,0}$,\,$\ket{0,1}$,\,$\ket{1,0}$,\,$\ket{1,1}$, with single-qubit rotations. Color images in Figs.\,1B-1E show the transient $\langle\VH\rangle$ as a function of  cavity drive frequency $\omega_{\mathrm{RF}}$ for the four cases (brackets denote an average over 600,000 repetitions).
On time scales $t \lesssim T_1^{\mathrm{L(R)}}=1.2(0.9)\, \us$, the largest transmission occurs at distinct frequencies for all four cases. The discrete transmission peaks are well resolved in Fig.\,1F, showing the time average of $\langle\VH\rangle$ versus frequency over the first $500\,\ns$. By matching these frequencies to Eq.\,(1), we extract $\chi^{\text{L(R)}}/2\pi=13(4)\,\MHz$. On long time scales  $t\gtrsim T_1^{\mathrm{L(R)}}$,
the dominant transmission is at the peak corresponding to $\ket{0,0}$ for all cases.  The transition between the two time limits is most evident  for $\ket{1,1}$, which decays partially into $\ket{0,1}$ and $\ket{1,0}$ before completely relaxing into the ground state $\ket{0,0}$.

It is because $\chiL$ and $\chiR$ are both larger than $\kappa= 1\, \MHz$ that a measurement can give joint qubit information. For example, applying a measurement tone at the cavity transmission peak corresponding to $\ket{0,0}$ queries for the joint property that \emph{both} qubits are in their ground state. Given the large state-dependent dispersive shifts, the system being in any other state results in a low transmission signal. This is best shown by the four transmission transients in Fig.\,1G. The transients for the states $\ket{0,1}$, $\ket{1,0}$ and $\ket{1,1}$ are all different from the transient for $\ket{0,0}$, but also not identical to each other. Qubit relaxation during the measurement results in the transients converging towards the $\ket{0,0}$ response on long time scales. 

More rigorously, the idealized measurement $M$ is a projection operator onto $\ket{0,0}$, $M=\ket{0,0}\bra{0,0}=(I+\ZL + \ZR+ \ZL \otimes \ZR)/4$, which is sensitive to the polarization of each qubit along its $z$-axis as well as to two-qubit correlations. However, the actual measurement is an ensemble average of $\VH$ which, due to qubit relaxation and partial overlap of the dispersive peaks, we expect to be described by  
\begin{equation}
    \langle \VH \rangle = \beta_{II}  + \beta_{ZI} \langle \ZL \rangle + \beta_{IZ} \langle \ZR \rangle + \beta_{ZZ} \langle \ZL \otimes \ZR \rangle+\delta v,
\end{equation}
where the $\beta_{LR}$ are constant coefficients and $\delta v$ is classical amplifier noise (details in supporting material). The amplifier noise, which limits the single-shot fidelity to $\sim$$10\%$, is largely eliminated by repeating the state preparation and measurement 600,000 times over fifteen seconds.

A simple set of Rabi flopping experiments can be performed (Figs.\,2A-C) to validate the measurement model of Eq.\,(2) and place bounds on other terms. The most general model, including all linear combinations of two-qubit Pauli operators \cite{otimes_note}, is
\begin{equation}
	M = \sum_{L,R \in \{I, X, Y, Z\}} \beta_{LR} L \otimes R.
\end{equation}
Figures 2A and 2B show the measured $\langle\VH\rangle$ as a function of the duration of an applied drive at $\wL$ and $\wR$, respectively. In each case the drive induces a Rabi oscillation of the addressed qubit around the $y$-axis of its Bloch sphere \cite{nielsen_chuang_2000}.  The observed oscillations in $\langle\VH\rangle$ are fit (solid lines) excellently by assuming the measurement model in Eq.\,(3) and the theoretical $\langle Z \rangle$ and $\langle X \rangle$ for driven qubits (see details in supporting material). From the fit to Fig.\,2A (2B), we can estimate deviations from Eq.\,(2) due to the terms $\langle\XL\rangle$ $(\langle\XR\rangle)$ and $\langle XZ\rangle$ $(\langle ZX \rangle)$ are each $\lesssim 2\%$ of the full range of $\langle \VH\rangle$.

\begin{figure}[ht!]
\centering
\includegraphics[scale=1.0]{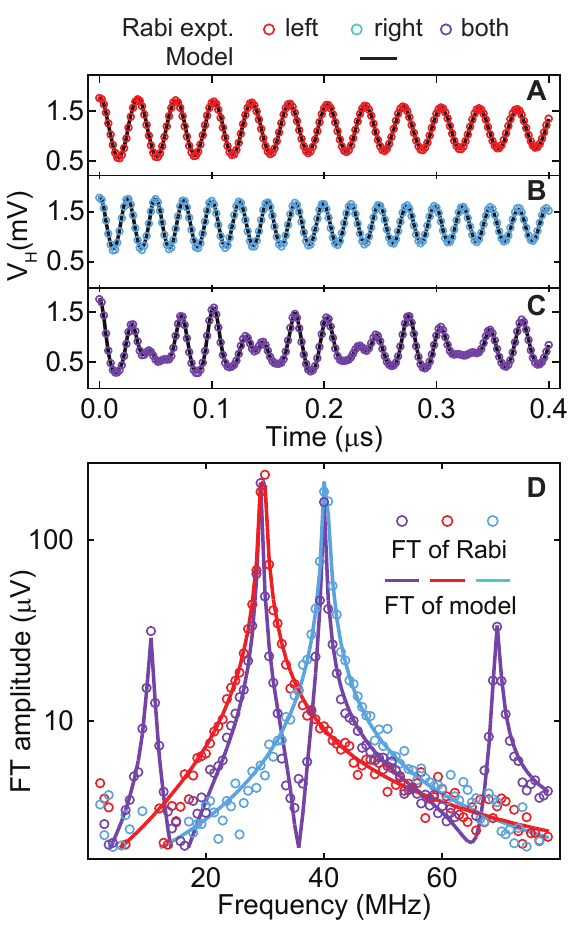}
\caption{Measurement model from Rabi oscillations. \textbf{(A-C)} Rabi oscillations on the \textbf{(A)} left qubit, \textbf{(B)} right qubit, and \textbf{(C)} simultaneously on both. Solid lines are fits to the model in Eq.\,(3). See text for details. \textbf{(D)} Fourier transform (FT) of the three Rabi experiments (markers) and of best fits (curves). While the red (blue) traces show one main peak at the Rabi frequency $\Omega^{\text{L(R)}}$, the purple traces reveal peaks at $\Omega^{\text{L}}$, $\Omega^{\text{R}}$, $\Omega^{\text{L}} +\Omega^{\text{R}}$, and $\Omega^{\text{L}}-\Omega^{\text{R}}$, demonstrating the mixing property that makes the joint measurement sensitive to qubit-qubit correlations.}
\end{figure}

The third experiment (Fig.\,2C) measures the homodyne response to simultaneous Rabi drives on both qubits. The observed oscillations not only show frequency components at the individual Rabi frequencies, $\Omega^{\text{L}}$ and $\Omega^{\text{R}}$, but also at their sum and difference. This is clearly revealed in the Fourier transform of the oscillations (Fig.~2D). This mixing effect agrees quantitatively with the term  $\langle ZZ \rangle$ in Eq.\,(2). An excellent fit is also obtained, giving deviations from Eq.\,(2) due to $\langle XX \rangle$ that are $\sim 2\%$ of the full swing of $\langle \VH \rangle$ (see details in the supporting material). These Rabi experiments thus corroborate the measurement model Eq.\,(2).  

Besides testing the measurement model, these Rabi-flopping experiments allow calibration of the coefficients $\beta_{LR}$ in Eq.\,(2). The best fits give $\beta_{II}=800\,\mu$V, $\beta_{IZ}=380\,\mu$V, $\beta_{ZI}=380\,\mu$V, and $\beta_{ZZ}=200\,\mu$V. The large ratios $\beta_{ZZ}/\beta_{IZ(ZI)}\approx 0.6$ indicate the high sensitivity of the joint readout to the qubit-qubit correlations.  

Having substantiated the physical nature and the quantitative model of the joint readout, we next use it to perform tomography of a variety of two-qubit states, both separable and entangled, produced on demand. We first perform an over-complete set of 30 raw measurements, each obtained by applying a different pair of simultaneous rotations of the qubits prior to measurement. The qubit rotations are chosen from the set $\{I, R_x^{\pm\pi}, R_x^{\pm\pi/2}, R_y^{\pm\pi/2}\}$ (see supporting material). Using the calibration of the $\beta_{LR}$ and least-squares linear estimation, we then construct the Pauli set $\Pauli$, whose 16 elements are the expectation values of the two-qubit Pauli operators, $\langle LR \rangle$, where $L, R \in \{I, X, Y, Z\}$. Two-qubit states can then be visualized by plotting the Pauli set in bar-graph format, in contrast to plotting the density-matrix in three-dimensional city-scape format~\cite{nielsen_chuang_2000}.

\begin{figure}[t!]
\centering
\includegraphics[scale=1.0]{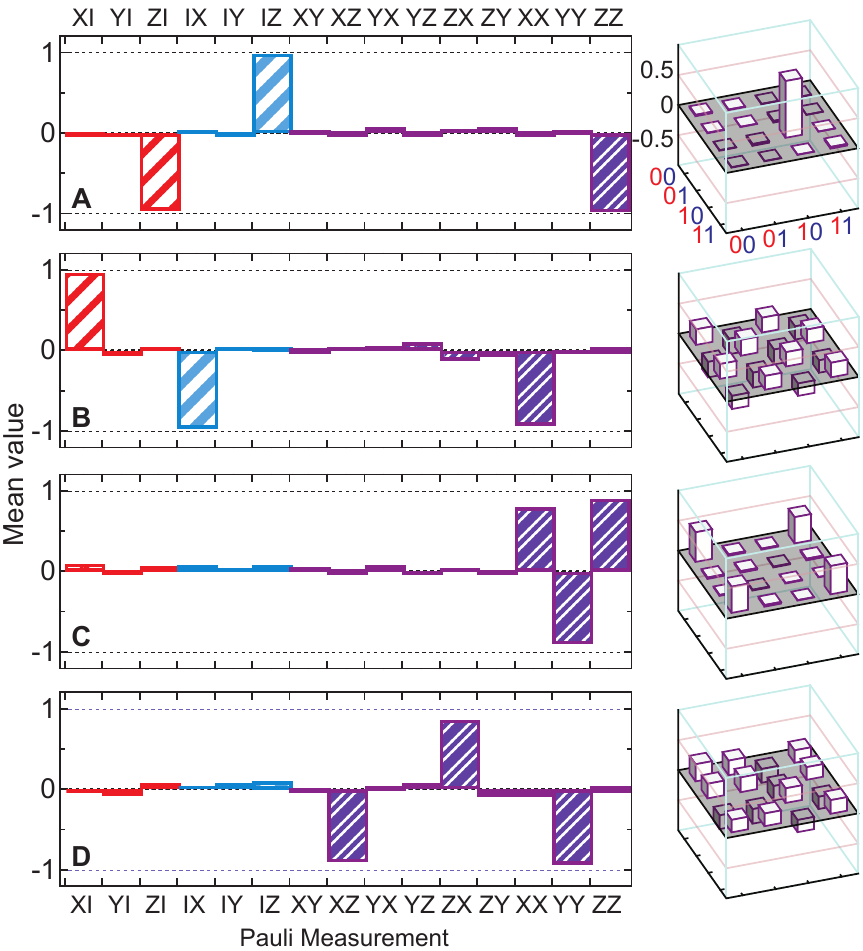}
\caption{Representation of two-qubit states using the Pauli set. Experimental Pauli set (with trivial $\langle II\rangle=1$ not shown), obtained from  linear operations on raw measurement data, for separable states \textbf{(A)} $\ket{1,0}$ and  \textbf{(B)} $(\ket{0,0}-\ket{0,1}+\ket{1,0}-\ket{1,1})/2$ and entangled states \textbf{(C)} $\ket{\Psi_+}$ and \textbf{(D)} the Bell state in the $x$-basis $\ket{\Phi_+}_x=(\ket{0,0}+\ket{0,1}-\ket{1,0}+\ket{1,1})/2$. Red (blue) bars correspond to left (right) single-qubit Pauli operators. Purple bars are the qubit-qubit correlations. The fidelities to the four ideal targeted states computed using Eq.\,(5) are $\F=98.2 \pm 0.4\%$, $96.8 \pm 0.4\%$, $90.0 \pm 0.6\%$, and $92.5 \pm 0.7\%$. The density matrix obtained using the same raw measurements is shown in three-dimensional city-scape format to the right of each Pauli set.}
\end{figure}

An advantage of the Pauli set representation is that one can easily distinguish separable from entangled states. In Fig.\,3, we use the Pauli set to view the experimentally generated separable states, (A) $\ket{0,1}$ and (B) $(\ket{0}+\ket{1})_{\text{L}}\otimes(\ket{0}-\ket{1})_\text{R}/2$, as well as entangled states, (C) $\ket{\Phi_+}$,  and (D) the Bell state in the $x$-basis  $\ket{\Psi_+}_x=(\ket{0,0}+\ket{1,0}-\ket{0,1}+\ket{1,1})/2$. For these four states, the Pauli set ideally contains three non-zero bars, all of unit magnitude. The Pauli set can be sub-divided into three sections:
the qubit polarization vectors, $\vec{P}^{\text{L}}=\{\langle\XL\rangle, \langle\YL\rangle,\langle\ZL\rangle\}$ and  $\vec{P}^{\text{R}}=\{\langle\XR\rangle, \langle\YR\rangle,\langle\ZR\rangle\}$, and the vector of two-qubit correlations $\vec{Q}=\{\langle XR \rangle, \cdots, \langle ZZ \rangle \}$. In the figure, $\vec{P}^{\text{L}}$, $\vec{P}^{\text{R}}$ and $\vec{Q}$ are color-coded
red, blue and purple, respectively. For the separable states, we observe near unity components in the three sub-sections of the Pauli set, $\vec{P}^{\text{L}}$, $\vec{P}^{\text{R}}$, and $\vec{Q}$. In contrast, for the entangled states, the only large components are in $\vec{Q}$. 
The presence of large bars in $\vec{Q}$ and vanishing $\vec{P}^{\text{L}}$, $\vec{P}^{\text{R}}$ is a direct signature of a high degree of entanglement.

The Pauli set representation permits testing of some simple physical constraints, which if not met, reveal systematic errors. The most easily tested physical constraint is $|\langle LR\rangle | \leq 1,\,\forall\, L,R$ (other constraints are $0\leq|\vec{P}^{\text{L,R}}|\leq 1$, and $0\leq|\vec{Q}|\leq \sqrt{3}$). Although the states shown in Fig.\,3 satisfy these bounds, a more thorough experimental test for systematic errors is to measure the Pauli set for a collection of states that differ only by the angle of a single-qubit rotation prior to measurement. Two such evolutions are shown in Figs.\,4A and 4B, which involve a rotation $\theta$ of the left qubit about its $y$-axis after preparing the separable state $\ket{0,0}$ (evolution A) and the entangled state $\ket{\Psi_+}_x$ (evolution B), respectively. 

In evolution A, varying $\theta$ produces the oscillation of the Pauli set shown in Fig.\,4C. Systematic errors in detection could appear as offsets and amplitudes of the Pauli set betraying the $\pm 1$ bounds. Such deviations would limit the accuracy of physical quantities extracted from the Pauli set, and thus are important to identify and correct. In Fig.\,4C, $\langle XI\rangle$, $\langle ZI\rangle$, $\langle XZ\rangle$, and $\langle ZZ\rangle$ oscillate with an average visibility of $97.6 \pm 0.3 \%$, demonstrating the large swing of the meter. Moreover, the average absolute error of all the ideally zero-valued bars is $\lesssim 10\%$. In evolution B, the dominant oscillating components are all in $\vec{Q}$, indicating that the state remains entangled throughout all the rotations. In this case, we find a visibility of $91.5\pm 0.3\%$, in good agreement with a master equation simulation taking into account qubit relaxation and dephasing. A residual oscillation amplitude of $\sim$$10\%$ is observed in $\langle XI\rangle$ and $\langle ZI \rangle$, a factor $\sim$$2$ larger than expected from theory. This discrepancy can arise from a combination of small calibration errors  in single-qubit rotations and various residual higher order couplings (see supporting material).  In the future, composite pulses and shaping might further reduce these effects.

\begin{figure*}[t!]
\centering
\includegraphics[scale=1.0]{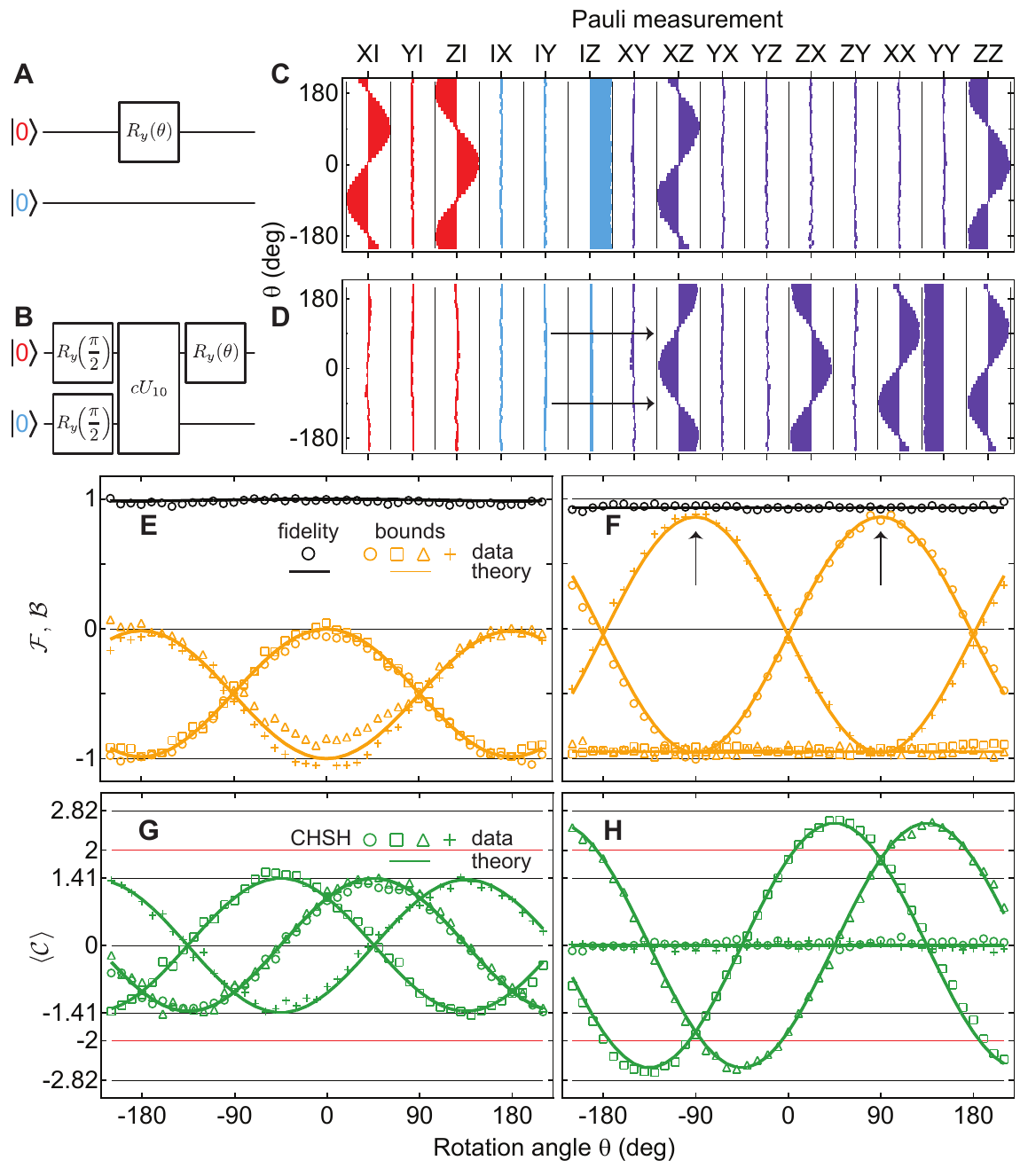}
\caption{Evolution of the Pauli set and entanglement witnesses under qubit rotations. (\textbf{A} and \textbf{B}) Gate
sequences generating evolutions of the Pauli set by subjecting (\textbf{A}) the separable state $\ket{0,0}$ and (\textbf{B}) the entangled state $\ket{\Psi_+}_x$ to  a rotation $R_y(\theta)$ on the left qubit, $-210^{\circ} \leq\theta \leq 210^{\circ}$. (\textbf{C} and \textbf{D}) The measured Pauli set as a function of $\theta$ for evolutions \textbf{A} and \textbf{B}, respectively. In evolution \textbf{A}, the left qubit polarization rotates along the $x$-$z$ plane, while the right qubit remains fully-polarized along $z$. In evolution \textbf{B}, both qubit polarizations vanish, with the only nonzero and oscillating Pauli operators being qubit-qubit correlators (purple bars). Arrows at $\theta=-90 (+90)^{\circ}$ indicate when the ideal two-qubit state is the standard Bell state $\ket{\Phi_-} (\ket{\Psi_+})$. (\textbf{E} and \textbf{F}) Experimental lower bounds $\mathcal{B}_i$ (orange) on the concurrence given by the optimal witnesses for Bell states $\mathcal{W}_{\Psi +} $\,(circles), $\mathcal{W}_{\Psi -}$\,(squares), $\mathcal{W}_{\Phi +}$\,(triangles), and $\mathcal{W}_{\Phi -}$\,(crosses), and fidelity $\mathcal{F}$ to the ideal state (black circles) for (\textbf{E}) evolution  \textbf{A} and (\textbf{F}) evolution \textbf{B}.  In \textbf{F}, a maximum lower bound is reached by $\mathcal{B}_{\Phi -}$ ($\mathcal{B}_{\Psi +}$) at $\theta=-90 (+90)^{\circ}$.  (\textbf{G-H}) Experimental average value of CHSH operators $\CHSHzzxx$ (circles), $\CHSHzxxz$ (squares), $\CHSHzxzx$ (triangles), $\CHSHxxzz$ (crosses). (\textbf{G}) For evolution \textbf{A} all $\mCHSH$ values stay within the separable state bounds $\pm \sqrt{2}$ up to measurement noise, while for (\textbf{H}) evolution \textbf{B}, $\max \amCHSH = 2.61 \pm 0.04$. Solid lines are master equation simulations.}
\end{figure*}

Finally, we use the fully characterized and high-visibility readout to extract accurate and precise measurements of physical quantities intrinsic to the two-qubit state. Examples include the fidelity to the targeted state, entanglement witnesses, and CHSH operators.  Interrogations of such quantities are henceforth statements about the quality of the states prepared.

The fidelity $\F$ to a targeted state provides one way of quantifying the control over two-qubit states and is given by the inner (dot) product of the measured $\Pauli$ to that of the ideal state, 
\begin{equation}
\F=\frac{1}{4} \Pauli \cdot \vec{P}_{\text{ideal}}.
\end{equation}
For the states produced in evolutions A and B, we find average fidelities $\mathcal{F}=98.8 \pm 1.0\%$ and $93.4 \pm 1.5\%$ over all $\theta$, respectively. The fidelities (black circles) in Figs.\,4E-F show excellent agreement with simulation (solid lines). This agreement demonstrates the accuracy of our meter for detecting both separable and entangled states, and the high fidelity values attest to the level of two-qubit control in the experiment.

Measures beyond fidelity are necessary to quantify the degree of two-qubit entanglement. Often, entanglement monotones such as concurrence $\conc$~\cite{horodecki_rmp} are obtained using non-linear estimators. It is standard to first perform maximum-likelihood estimation \cite{James_qubitmeas_2001} to generate a physical $\rho$ despite any statistical or systematic errors in the raw data, and to then calculate these metrics from the eigenvalue spectrum of related matrices \cite{horodecki_rmp}.  This non-linear process complicates the propagation of any statistical and systematic errors in the measurements, and can bias the estimation of such metrics as the purity of the two-qubit state increases \cite{blume,Home:2009fv} (See supporting material).
 
To be quantitative about entanglement while using only linear operations on the raw measurements, we make use of entanglement witnesses \cite{horodecki_rmp,eisert_witness_2007}. An entanglement witness is a unity-trace observable $\W$ with a positive expectation value for all separable states, such that $\tr(\rho \W)<0$ guarantees entanglement. Furthermore, $-2 \tr(\rho \W)$ gives a lower bound on $\conc$~\cite{eisert_witness_2007}. The optimal witness for a targeted entangled state gives the tightest lower bound. For the Bell states, these are
\begin{eqnarray*}
	\mathcal{W}_{\Psi +} &=& \frac{1}{4} (II - XX + YY - ZZ), \\
	\mathcal{W}_{\Psi -} &=& \frac{1}{4} (II+XX - YY-ZZ ),\\
	\mathcal{W}_ {\Phi +} &=& \frac{1}{4} (II-XX-YY+ZZ), \\
	\mathcal{W}_{\Phi -} &=& \frac{1}{4}(II+XX+YY+ZZ),
\end{eqnarray*}
giving lower bounds $\mathcal{B}_i = -2 \tr(\rho \mathcal{W}_i)$ on $\conc$.

The four bounds obtained for evolution A(B) are shown in Fig.\,4E(4F). In evolution A, the four bounds are non-positive for all $\theta$ to within measurement error, indicating that entanglement is not witnessed. This is as expected, since single-qubit rotations should not produce any entanglement. In Fig.\,4F, in contrast, bounds $\B_{\Psi +}$ and $\B_{\Phi -}$ extend into the positive region, reaching $85.9\pm 1.5\%$ and $88.1 \pm 1.5\%$ at $\theta=-90^{\circ}$ and $90^{\circ}$, respectively. There is at least one positive bound for most $\theta$ (excluding $\pm 180^{\circ}$ and $0^{\circ}$), indicating that the two qubits are entangled. Solid lines in Figs.\,4E-F are obtained from the master equation simulation. The agreement between data and theory shows the accuracy of the entanglement witnesses and the small residuals $\sim$$2\%$ demonstrate their precision. 

A more well-known entanglement measure is the CHSH operator, often used to test local-hidden variable theories. CHSH operators \cite{clauser_proposed_1969} are most generally defined as 
\begin{eqnarray}
	\CHSH_{A,B,A',B'}= AB +A B' + A'B -  A'B',
\end{eqnarray}
with left-qubit operators $A$, $A'$ and right-qubit operators $B$, $B'$ being single-qubit Pauli operators along any axis. For separable states, $\amCHSH \leq 2$. However, for $A \perp A'$ and $B \perp B'$, the separable bound is tighter, $\amCHSH \leq \sqrt{2}$.  
 
From the Pauli sets of evolutions A and B, we obtain experimental expectation values of four CHSH operators with $A,A'\in \{\XL,\ZL\}$ and $B,B'\in \{\XR,\ZR\}$. For the separable states of evolution A (Fig.\,4G), we find that to within statistical error, $\mCHSHzxxz=\mXX-\mXZ+\mZX+\mZZ$ (squares) and $\mCHSHzxzx=\mXX+\mXZ-\mZX+\mZZ$ (triangles) remain within the separable bounds for all $\theta$. For the entangled states prepared in evolution B (Fig.\,4H), instead, $\mCHSH$ oscillates past the separable bounds. At $\theta = \pm 45^{\circ}$, a maximum value $\amCHSH=2.61 \pm 0.04$ is reached. The agreement with theory and proximity of the maximum $\amCHSH$ to the $2\sqrt{2}$ upper bound \cite{Cirelson_1980} further demonstrate the high degree of entanglement of the states produced. We note that while $\mCHSH$ exceeds the local-hidden variable bound \cite{horodecki_rmp} of 2 by $\sim$$15$ standard deviations, this result is not a strict violation of local-hidden variable theories due to locality and measurement loopholes in our system. For superconducting qubits, the measurement loophole has recently been closed using high-fidelity single-shot readout \cite{ansmann_thesis}. 

In summary, we have  performed metrology of entanglement between two superconducting qubits using a single measurement channel giving direct access to qubit correlations. While loopholes  preclude using this joint readout for fundamental tests of quantum mechanics, the demonstrated ability to measure fidelity to targeted states, lower bounds on concurrence and strong violation of CHSH inequalities is eminently useful for quantum-computer engineering. In its present form, this single-channel detector will be able to detect future improvements in the controlled production of entangled states and is immediately extendable to three or four qubits. Similar implementations of joint detection may prove useful for metrology of multi-qubit operations in other qubit platforms  where cross-talk and low measurement fidelity make single-shot individual qubit readouts challenging.\\
\begin{acknowledgements}
	We acknowledge discussions with R.\,Blume-Kohout, B.\,R.\,Johnson, Jens Koch, N.\,L\"utkenhaus,  K.\,Resch, C.\,Rigetti, and D.\,I.\,Schuster. This work was supported by LPS/NSA under ARO Contract No.\,W911NF-05-1-0365,
and by the NSF under Grants No.\,DMR-0653377 and No.\,DMR-0603369.
We acknowledge additional support from a CIFAR Junior Fellowship, MITACS, MRI, and NSERC (JMG), and from CNR-Istituto di Cibernetica, Pozzuoli, Italy (LF).
\end{acknowledgements}


\begin{thebibliography}{10}
\newcommand{\enquote}[1]{``#1''}

\bibitem{bell_einstein-podolsky-rosen_1964}
J.~S. Bell, Physics \textbf{1}, 195 (1964).

\bibitem{EPR}
A.~Einstein, B.~Podolsky, and N.~Rosen, Phys. Rev. \textbf{47}, 777 (1935).

\bibitem{CH_bell}
J.~F. Clauser and M.~A. Horne, Phys. Rev. D \textbf{10}, 526 (1974).

\bibitem{aspect_bells_1982}
A.~Aspect, J.~Dalibard, and G.~Roger, Phys. Rev. Lett. \textbf{49}, 1804
  (1982).

\bibitem{tittel_bells_1998}
W.~Tittel, J.~Brendel, H.~Zbinden, and N.~Gisin, Phys. Rev. Lett. \textbf{81},
  3563 (1998).

\bibitem{zeilinger_experiment_1999}
A.~Zeilinger, Rev. Mod. Phys. \textbf{71}, 288 (1999).

\bibitem{ansmann_thesis}
M.~Ansmann, Ph.D. thesis, University of California at Santa Barbara (2009).

\bibitem{horodecki_rmp}
R.~Horodecki, P.~Horodecki, M.~Horodecki, and K.~Horodecki, Rev. Mod. Phys.
  \textbf{81}, 865 (2009).

\bibitem{Haffner_multiparticle_ions_2005}
H.~Haffner, W.~Hansel, C.~F. Roos, J.~Benhelm, D.~Chek-al kar, M.~Chwalla,
  T.~Korber, U.~D. Rapol, M.~Riebe, P.~O. Schmidt, C.~Becher, O.~Guhne, W.~Dur,
  and R.~Blatt, Nature \textbf{438}, 643 (2005).

\bibitem{Leibfried:2005yq}
D.~Leibfried, E.~Knill, S.~Seidelin, J.~Britton, R.~B. Blakestad,
  J.~Chiaverini, D.~B. Hume, W.~M. Itano, J.~D. Jost, C.~Langer, R.~Ozeri,
  R.~Reichle, and D.~J. Wineland, Nature \textbf{438}, 639 (2005).

\bibitem{myerson:200502}
A.~H. Myerson, D.~J. Szwer, S.~C. Webster, D.~T.~C. Allcock, M.~J. Curtis,
  G.~Imreh, J.~A. Sherman, D.~N. Stacey, A.~M. Steane, and D.~M. Lucas, Phys.
  Rev. Lett. \textbf{100}, 200502 (2008).

\bibitem{martinis_rabi_2002}
J.~M. Martinis, S.~Nam, J.~Aumentado, and C.~Urbina, Phys. Rev. Lett.
  \textbf{89}, 117901 (2002).

\bibitem{siddiqi_jba_2004}
I.~Siddiqi, R.~Vijay, F.~Pierre, C.~M. Wilson, M.~Metcalfe, C.~Rigetti,
  L.~Frunzio, and M.~H. Devoret, Phys. Rev. Lett. \textbf{93}, 207002 (2004).

\bibitem{Lupascu:2007nx}
A.~Lupascu, S.~Saito, T.~Picot, P.~C. de~Groot, C.~J. P.~M. Harmans, and J.~E.
  Mooij, Nature Phys. \textbf{3}, 119 (2007).

\bibitem{trauzettel:235331}
B.~Trauzettel, A.~N. Jordan, C.~W.~J. Beenakker, and M.~B{\"u}ttiker, Phys.
  Rev. B \textbf{73}, 235331 (2006).

\bibitem{Mao_mesoQM_2004}
W.~Mao, D.~V. Averin, R.~Ruskov, and A.~N. Korotkov, Phys. Rev. Lett.
  \textbf{93}, 056803 (2004).

\bibitem{wallra_strong_2004}
A.~Wallraff, D.~I. Schuster, A.~Blais, L.~Frunzio, R.-S. H.~J. Majer, S.~Kumar,
  S.~M. Girvin, and R.~J. Schoelkopf, Nature \textbf{431}, 162 (2004).

\bibitem{filipp_joint_2009}
S.~Filipp, P.~Maurer, P.~J. Leek, M.~Baur, R.~Bianchetti, J.~M. Fink, M.~Goppl,
  L.~Steffen, J.~M. Gambetta, A.~Blais, and A.~Wallraff, Phys. Rev. Lett.
  \textbf{102}, 200402 (2009).

\bibitem{clauser_proposed_1969}
J.~F. Clauser, M.~Horne, A.~Shimony, and R.~A. Holt, Phys. Rev. Lett.
  \textbf{23}, 880 (1969).

\bibitem{koch_charge-insensitive_2007}
J.~Koch, T.~M. Yu, J.~Gambetta, A.~A. Houck, D.~I. Schuster, J.~Majer,
  A.~Blais, M.~H. Devoret, S.~M. Girvin, and R.~J. Schoelkopf, Phys. Rev. A
  \textbf{76}, 042319 (2007).

\bibitem{schreier_suppressing_2008}
J.~A. Schreier, A.~A. Houck, J.~Koch, D.~I. Schuster, B.~R. Johnson, J.~M.
  Chow, J.~M. Gambetta, J.~Majer, L.~Frunzio, M.~H. Devoret, S.~M. Girvin, and
  R.~J. Schoelkopf, Phys. Rev. B \textbf{77}, 180502 (2008).

\bibitem{chow_bm_2009}
J.~M. Chow, J.~M. Gambetta, L.~Tornberg, J.~Koch, L.~S. Bishop, A.~A. Houck,
  B.~R. Johnson, L.~Frunzio, S.~M. Girvin, and R.~J. Schoelkopf, Phys. Rev.
  Lett. \textbf{102}, 090502 (2009).

\bibitem{motzoi_2009}
F.~Motzoi, J.~M. Gambetta, P.~Rebentrost, and F.~K. Wilhelm, arXiv:0901.0534
  (2009).

\bibitem{dicarlo_2009}
L.~DiCarlo, J.~M. Chow, J.~M. Gambetta, L.~S. Bishop, B.~R. Johnson, D.~I.
  Schuster, J.~Majer, A.~Blais, L.~Frunzio, S.~M. Girvin, and R.~J. Schoelkopf,
  Nature \textbf{460}, 240 (2009).

\bibitem{nielsen_chuang_2000}
M.~A. Nielsen and I.~L. Chuang, \emph{Quantum Computation and Quantum
  Information} (Cambridge University Press, Cambridge, 2000).

\bibitem{chow_supp}
 Materials and methods available in supporting material.

\bibitem{schuster_resolving_2007}
D.~I. Schuster, A.~A. Houck, J.~A. Schreier, A.~Wallraff, J.~M. Gambetta,
  A.~Blais, L.~Frunzio, B.~Johnson, M.~H. Devoret, S.~M. Girvin, and R.~J.
  Schoelkopf, Nature \textbf{445}, 515 (2007).

\bibitem{otimes_note}
 For the rest of the text we adopt the notation ${L} \otimes {R} \to
  {LR}$ for ${L},{R}\in\{{I},{X},{Y},{Z}\}$.

\bibitem{James_qubitmeas_2001}
D.~F.~V. James, P.~G. Kwiat, W.~J. Munro, and A.~G. White, Phys. Rev. A
  \textbf{64}, 052312 (2001).

\bibitem{blume}
R.~Blume-Kohout, arXiv:quant-ph/0611080v1  (2006).

\bibitem{Home:2009fv}
J.~P. Home, D.~Hanneke, J.~D. Jost, J.~M. Amini, D.~Leibfried, and D.~J.
  Wineland, Science, DOI: 10.1126/science.1177077 (6 August 2009).

\bibitem{eisert_witness_2007}
J.~Eisert, F.~G. S.~L. Brandao, and K.~M.~R. Audenaert, New J. Phys.
  \textbf{9}, 46 (2007).

\bibitem{Cirelson_1980}
B.~S. Cirel'son, Lett. Math. Phys. \textbf{4}, 93 (1980).

\end{thebibliography}
\end{document}


\title{Supporting Material for ``Entanglement Metrology Using a Joint Readout of Superconducting Qubits''}
\date{\today}
\author{J. M. Chow}
\affiliation{Departments of Physics and Applied Physics, Yale University, New Haven, Connecticut 06520, USA}
\author{L. DiCarlo}
\affiliation{Departments of Physics and Applied Physics, Yale University, New Haven, Connecticut 06520, USA}
\author{J. M. Gambetta}
\affiliation{Institute for Quantum Computing and Department of Physics and Astronomy, University of Waterloo, Waterloo, Ontario, Canada N2L 3G1}
\author{A. Nunnenkamp}
\affiliation{Departments of Physics and Applied Physics, Yale University, New Haven, Connecticut 06520, USA}
\author{Lev S. Bishop}
\affiliation{Departments of Physics and Applied Physics, Yale University, New Haven, Connecticut 06520, USA}
\author{L. Frunzio}
\affiliation{Departments of Physics and Applied Physics, Yale University, New Haven, Connecticut 06520, USA}
\author{M. H. Devoret}
\affiliation{Departments of Physics and Applied Physics, Yale University, New Haven, Connecticut 06520, USA}
\author{S. M. Girvin}
\affiliation{Departments of Physics and Applied Physics, Yale University, New Haven, Connecticut 06520, USA}
\author{R. J. Schoelkopf}
\affiliation{Departments of Physics and Applied Physics, Yale University, New Haven, Connecticut 06520, USA}

\maketitle

\section{I. Materials and Methods}

\subsection{A. Sample Fabrication}
The device is fabricated on a 430 $\mu$m thick sapphire substrate. The superconducting transmission-line cavity and flux-bias lines are defined via optical lithography and fluorine-based reactive ion etching of a dc-sputtered niobium film (180 nm thick). The two transmons are patterned using electron-beam lithography with split-junctions, grown using double-angle deposition of aluminum, with layer thicknesses of 20 nm and 90 nm. The sample is cooled in a dilution refrigerator to $13~\mK$.
The experimental setup is presented in detail in a previous publication \cite{dicarlo_2009}.

\subsection{B. Experimentally Determined Parameters}
From a set of heterodyne transmission measurements obtained when tuning each qubit into near resonance with the cavity, we determine the qubit-cavity coupling strengths $g_{\text{L(R)}}/2\pi=199\,(183)$ MHz. Fitting a multi-level Tavis-Cummings Hamitonian to spectroscopy measurements of the two lowest transitions of each qubit, we extract maximum Josephson energies $E_{\text{J,L(R)}}^{\text{max}}/h=28.48\,(42.34)$ GHz and electrostatic charging energies $E_{\text{C,L(R)}}/h=317\,(297)$ MHz.
Standard sliding $\pi$-pulse and Ramsey fringe experiments give qubit relaxation times $T_1^{\text{L(R)}}=1.2\,(0.9)\,\us$ and dephasing times $T_2^{\text{L(R)}}=1.5\,(1.1)\,\us$.

\section{II. Experimental Reduction of Higher Level Leakage}

For superconducting qubits, previous work \cite{lucero_gates_2008, chow_bm_2009} has demonstrated single-qubit gate errors of 1-2\%. Specifically, in Ref.\,\cite{chow_bm_2009}, this average gate error is consistent with qubit relaxation and dephasing for gate lengths $t_{\text{gate}} > 16$ ns, and with leakage to the second excited level at shorter gate lengths. Recent theory proposes a possible reduction of errors due to this leakage to the second excited level via the technique of derivative removal via adiabatic gate (DRAG) \cite{motzoi_2009}. In DRAG, when a single-qubit rotation pulse is applied along the $x$-axis, a derivative of the pulse is applied along the $y$-axis to cancel out the higher level leakage, and vice versa. 

In this work, we have used DRAG on all single-qubit gates. Standard $x$- and $y$-rotations are performed with in-phase and quadrature microwaves tuned to the qubit ground to first excited state frequency, and are shaped with Gaussian envelopes, truncated to two standard deviations $\sigma$ on each side. After each gate, a $5\,\ns$ buffer is included to avoid any overlap with the
following gate. The leakage due to an $x\,(y)$-rotation or in-phase (quadrature) pulse is reduced by applying a complementary quadrature (in-phase) tone shaped by a truncated derivative-of-Gaussian envelope.

\begin{figure}[t!]
\centering
\includegraphics[scale=1.0]{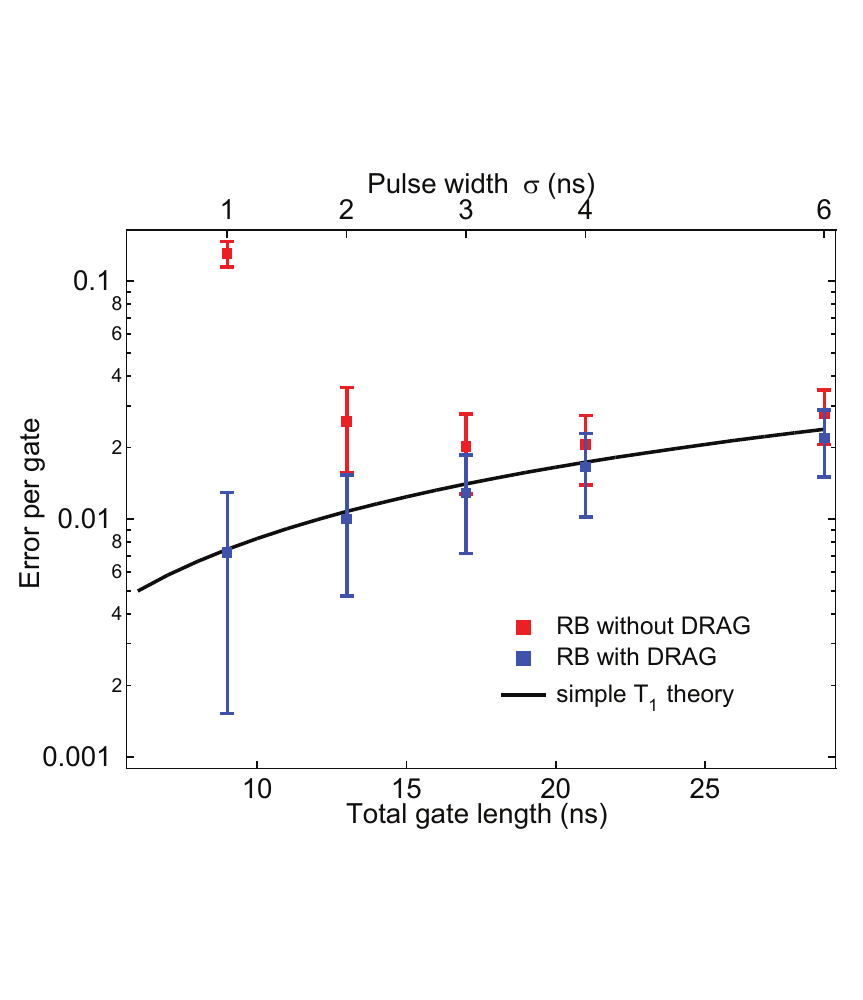}
\caption{\textbf{Reduction of single-qubit gate errors via DRAG.} Error per gate for the left qubit extracted from randomized benchmarking for different gate lengths using both standard Gaussian-shaped pulses (red squares) and DRAG-enhanced pulses (blue squares). Excellent overlap with a model for gate error including qubit relaxation (black curve) suggests that DRAG-enhanced pulses
successfully eliminate leakage to the second-excited state. Gate errors $\sim 1\%$, which are otherwise unattainable with standard pulses, are reached using DRAG.\label{figureS1}}
\end{figure}

Using the technique of randomized benchmarking (RB) \cite{knill_randomized_2008}, we extend the extraction of gate errors as performed in \cite{chow_bm_2009} to our single-qubit gates with and without DRAG, finding a significant improvement in the error rates with DRAG. As shown in Fig.\,S1, we find that without DRAG (red circles), the gate errors decrease with decreasing gate length before turning back up at a gate length of 17 ns, signaling the onset of second-excitation leakage. The overall higher and earlier onset of the increase in gate error compared with Ref.\,\cite{chow_bm_2009} is due in part to the shorter $T_1$ and the larger $g$ of the present qubit. In contrast, with DRAG (blue circles), the gate errors continue to decrease, reaching $0.7\pm 0.5\%$ at the shortest gate lengths possible with our arbitrary waveform generator (Tektronix 5014B).

The total gate length chosen for this work is 13 ns ($\sigma=2$ ns) at which the error per gate is  $0.9\pm 0.5\%$, in excellent agreement with the simple theory (solid black line) with $T_1 = 1.2 \text{ }\mu\text{s}$ for the left qubit. In the future, using DRAG with improved pulse-shaping resolution in hardware and improved coherence times could further lower single-qubit error rates towards the fault tolerant threshold \cite{Gottesman}.

\section{III. Measurement Model Validation}

As mentioned in the main text, we use fits to the three Rabi-flopping experiments of Fig.\,2 to place bounds on deviations from the measurement model of Eq.\,(2). Because in these tests each qubit is driven around the $y$-axis of its Bloch sphere, all terms involving $\YL$ and $\YR$ in Eq.\,3 would not contribute to $\mVH$. The presence of such terms can be tested by rotating each or both qubits around their $x$-axis instead. We do not find any significant differences in such experiments from the ones presented in the text, and the results here can be generalized for both quadratures $X$ and $Y$.

In  our experiment the detuning $\sim$$1.5\text{ GHz}$ between the two qubits is large compared to the Rabi-flopping rates, and we can assume a simple model of independent qubit driving. For a qubit driven at a rate $\Omega$ around its $y$-axis starting from the ground sate, the theoretical time evolution of $\langle Z \rangle$ and $\langle X\rangle$ is given by
\begin{align*}
\langle Z\rangle (t) &= \frac{\gamma_1\gamma_2}{\gamma_1\gamma_2 +\Omega^2} +
\frac{e^{-t/\tau_R}\Omega^2}{\gamma_1\gamma_2
   +\Omega^2}\left(\cos(\tilde{\Omega}t)+ \frac{\sin(\tilde{\Omega}
t)}{\tau_R\tilde{\Omega}} \right),&& \\
\langle X\rangle (t) &= \frac{\gamma_1 \Omega}{\gamma_1 \gamma_2+
\Omega^2}-\frac{e^{-t/\tau_R}\Omega}{\gamma_1\gamma_2+\Omega^2}&&\\
   &&\mathllap{\times\left(\gamma_1\cos(\tilde{\Omega}t)-
           \frac{\big[2\Omega^2+\gamma_1(\gamma_2-\gamma_1)\big]\sin(\tilde{\Omega}t)}{2\tilde{\Omega}}\right).}&\\
\end{align*}
 Here, $\tilde{\Omega}= \sqrt{\Omega^2-(1/\tau_R)^2}$ is an effective oscillation rate,
$\gamma_1=1/T_1$ is the relaxation rate, $\gamma_2=\gamma_1/2+\gamma_{\phi}$ is the dephasing rate, and
$ \tau_R = 2/(\gamma_1+\gamma_2)$ is the Rabi decay time.

 Using these expressions in Eq.~3 and fitting to the three experiments, we can estimate the coefficients $\beta_{LR}$. For single-qubit driving [Fig.\,2A(B)], the right (left) qubit is always in the ground state, and only terms
  $\mZL$, $\mXL$,  $\mXZ$ and $\mZZ$ ($\mZR$, $\mXR$, $\mZX$, and  $\mZZ$)  contribute to the $\mVH$ oscillation. Fitting the form
 \begin{eqnarray*}
\mVH_{A(B)} = W_0 + W_1 \langle Z^{\text{L(R)}} \rangle + W_2 \langle X^{\text{L(R)}}\rangle,
\end{eqnarray*}
with $W_0$, $W_1$, $W_2$, $\Omega^{\text{L(R)}}$, $\gamma_1^{\text{L(R)}}$, and $\gamma_2^{\text{L(R)}}$ as free parameters gives an
excellent fit. In both cases, the best-fit $W_2$, corresponding to $\beta_{XI(IX)} + \beta_{XZ(ZX)}$, is less than $2\%$ of the full range of $\mVH$, $\sim 2 \beta_{IZ}+2\beta_{ZI}$. For the doubly-driven case (Fig.~2C), the fit function used is
\begin{eqnarray*}
\langle V_H\rangle_{C}=&\beta_{II}+\beta_{XI} \langle \XL \rangle + \beta_{ZI} \langle \ZL \rangle + \beta_{IX} \langle \XR \rangle +\beta_{IZ} \langle \ZR\rangle \\
&+ \beta_{XX} \langle XX\rangle + \beta_{XZ} \langle XZ \rangle + \beta_{ZX} \langle ZX \rangle + \beta_{ZZ} \langle ZZ \rangle ,
\end{eqnarray*}
with $\beta_{ij}$, $\Omega^{\text{L}}$, $\Omega^{\text{R}}$, $\gamma_j^{\text{L}}$, and $\gamma_j^{\text{R}}$  as fit parameters. The best-fit coefficients captured in Eq.\,(2) are $\beta_{II}=800\,\mu$V, $\beta_{IZ} = 380\,\mu$V, $\beta_{ZI}=380\,\mu$V, $\beta_{ZZ} = 200\,\mu$V. Best-fit values of the remaining coefficients are each less than $2\%$ of the full range of $\mVH$.

\section{IV. Two-Qubit State Tomography}

Full tomography of the two-qubit state is performed by using an over-complete set of 30 raw measurements. These measurements involve applying different simultaneous rotations on the qubits, as listed in Table S1. The 15 measurements labeled $P_i$ involve positive rotations chosen from $\{I, R_x^{+\pi}, R_x^{+\pi/2}, R_y^{+\pi/2}\}$. The remaining 15, labeled $N_i$, involve negative rotations chosen from $\{I, R_x^{-\pi}, R_x^{-\pi/2}, R_y^{-\pi/2}\}$. Ensemble averages of $P_i$ and $N_i$ are obtained by repeating state preparation, analysis rotation, and measurement over 600,000 times. A standard least-squares linear estimator is then used to extract the Pauli set $\vec{P}$ discussed in the text. Although just  15  linearly independent measurements (such as either all $P_i$ or all $N_i$) is necessary for state tomography, using all of these rotations and least-squares estimation reduces the statistical error in the extraction of $\vec{P}$. 

\begin{table}[h!]
\centering
\begin{tabular}{c|rcl|rrr}
\hline\hline
Msmt. &\multicolumn{3}{c|}{Pre-rotation} & \multicolumn{3}{c}{Ensemble average}\\
\hline
$P_{01}$& $\Id$ &$\otimes$& $\Id$     &   $+\bL\mZL$&$+\bR\mZR$&$+\bLR\mZZ$\\
$P_{02}$& $\Rxp$ &$\otimes$& $\Id$    &   $-\bL\mZL$&$+\bR\mZR$&$-\bLR\mZZ$\\
$P_{03}$& $\Id$ &$\otimes$& $\Rxp$    &   $+\bL\mZL$&$-\bR\mZR$&$-\bLR\mZZ$\\
$P_{04}$& $\Rxpt$ &$\otimes$& $\Id$   &   $+\bL\mYL$&$+\bR\mZR$&$+\bLR\mYZ$\\
$P_{05}$& $\Rxpt$ &$\otimes$& $\Rxpt$ &   $+\bL\mYL$&$+\bR\mYR$&$+\bLR\mYY$\\
$P_{06}$& $\Rxpt$ &$\otimes$& $\Rypt$ &   $+\bL\mYL$&$-\bR\mXR$&$-\bLR\mYX$\\
$P_{07}$& $\Rxpt$ &$\otimes$& $\Rxp$  &   $+\bL\mYL$&$-\bR\mZR$&$-\bLR\mYZ$\\
$P_{08}$& $\Rypt$ &$\otimes$& $\Id$   &   $-\bL\mXL$&$+\bR\mZR$&$-\bLR\mXZ$\\
$P_{09}$& $\Rypt$ &$\otimes$& $\Rxpt$ &   $-\bL\mXL$&$+\bR\mYR$&$-\bLR\mXY$\\
$P_{10}$& $\Rypt$ &$\otimes$& $\Rypt$ &  $-\bL\mXL$&$-\bR\mXR$&$+\bLR\mXX$\\
$P_{11}$& $\Rypt$ &$\otimes$& $\Rxp$  &  $-\bL\mXL$&$-\bR\mZR$&$+\bLR\mXZ$\\
$P_{12}$& $\Id$ &$\otimes$& $\Rxpt$   &  $+\bL\mZL$&$+\bR\mYR$&$+\bLR\mZY$\\
$P_{13}$& $\Rxp$ &$\otimes$& $\Rxpt$  &  $-\bL\mZL$&$+\bR\mYR$&$-\bLR\mZY$\\
$P_{14}$& $\Id$ &$\otimes$& $\Rypt$   &  $+\bL\mZL$&$-\bR\mXR$&$-\bLR\mZX$\\
$P_{15}$& $\Rxp$ &$\otimes$& $\Rypt$  &  $-\bL\mZL$&$-\bR\mXR$&$+\bLR\mZX$\\

$N_{01}$& $\Id$ &$\otimes$& $\Id$     &   $+\bL\mZL$&$+\bR\mZR$&$+\bLR\mZZ$\\
$N_{02}$& $\Rxm$ &$\otimes$& $\Id$    &   $-\bL\mZL$&$+\bR\mZR$&$-\bLR\mZZ$\\
$N_{03}$& $\Id$ &$\otimes$& $\Rxm$    &   $+\bL\mZL$&$-\bR\mZR$&$-\bLR\mZZ$\\
$N_{04}$& $\Rxmt$ &$\otimes$& $\Id$   &   $-\bL\mYL$&$+\bR\mZR$&$-\bLR\mYZ$\\
$N_{05}$& $\Rxmt$ &$\otimes$& $\Rxmt$ &   $-\bL\mYL$&$-\bR\mYR$&$+\bLR\mYY$\\
$N_{06}$& $\Rxmt$ &$\otimes$& $\Rymt$ &   $-\bL\mYL$&$+\bR\mXR$&$-\bLR\mYX$\\
$N_{07}$& $\Rxmt$ &$\otimes$& $\Rxm$  &   $-\bL\mYL$&$-\bR\mZR$&$+\bLR\mYZ$\\
$N_{08}$& $\Rymt$ &$\otimes$& $\Id$   &   $+\bL\mXL$&$+\bR\mZR$&$+\bLR\mXZ$\\
$N_{09}$& $\Rymt$ &$\otimes$& $\Rxmt$ &   $+\bL\mXL$&$-\bR\mYR$&$-\bLR\mXY$\\
$N_{10}$& $\Rymt$ &$\otimes$& $\Rymt$ &  $+\bL\mXL$&$+\bR\mXR$&$+\bLR\mXX$\\
$N_{11}$& $\Rymt$ &$\otimes$& $\Rxm$  &  $+\bL\mXL$&$-\bR\mZR$&$-\bLR\mXZ$\\
$N_{12}$& $\Id$ &$\otimes$& $\Rxmt$   &  $+\bL\mZL$&$-\bR\mYR$&$-\bLR\mZY$\\
$N_{13}$& $\Rxm$ &$\otimes$& $\Rxmt$  &  $-\bL\mZL$&$-\bR\mYR$&$+\bLR\mZY$\\
$N_{14}$& $\Id$ &$\otimes$& $\Rymt$   &  $+\bL\mZL$&$+\bR\mXR$&$+\bLR\mZX$\\
$N_{15}$& $\Rxm$ &$\otimes$& $\Rymt$  &  $-\bL\mZL$&$+\bR\mXR$&$-\bLR\mZX$\\
\hline\hline
\end{tabular}
\caption{{\bfseries The 30 raw measurements.}
\label{tab:tomography}}
\end{table}

\section{V. Biasing of Metrics by Maximum-Likelihood Estimation}
Maximum-likelihood estimators (MLE) can become biased if the true mean lies close to a boundary of the allowed parameter space. In order to quantify the importance of this effect on the estimation of lower bounds on concurrence $\conc$ given by entanglement witnesses, we have performed Monte-Carlo simulations for nearly-pure Werner states \cite{horodecki_rmp},
\begin{equation*}
\rho_{W}(\lambda)=\lambda \ket{\psi^-}\bra{\psi^-} + (1-\lambda) I/4,
\end{equation*}
with Werner parameter $\lambda \in \left[0.8,1\right]$.
\begin{figure}[t!]
\centering
\includegraphics[width=3.4 in]{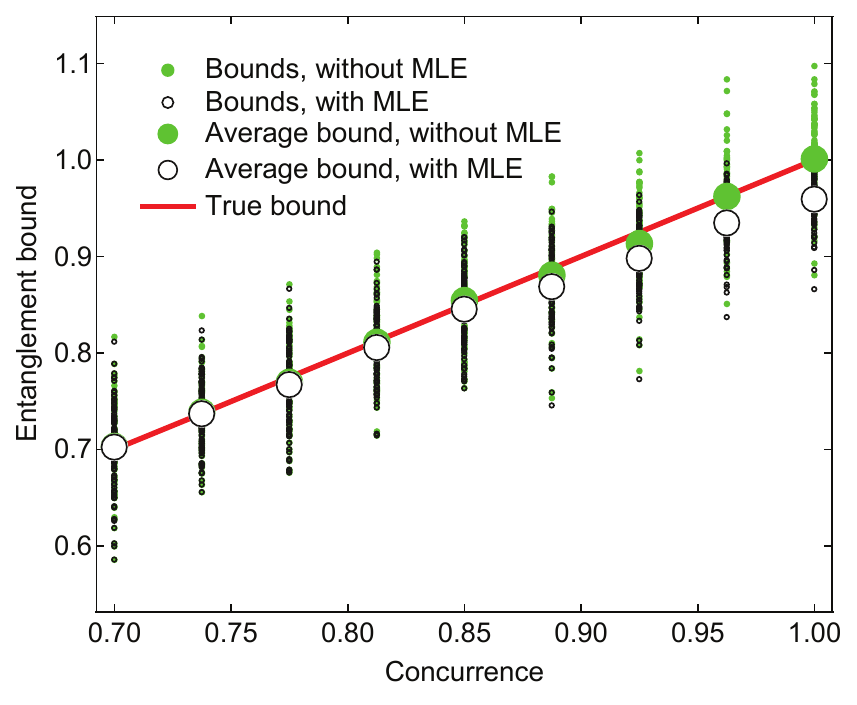}
\caption{\textbf{Biasing of entanglement bounds by MLE.}  Comparison of the lower bound $\B_{\psi -}$ on concurrence $\conc$ computed from simulated noisy raw data with and without use of maximum-likelihood estimation. The bound computed using MLE systematically underestimates the true bound (in this case, always equal to the true concurrence $\conc$, red line), while the bound computed directly from the simulated raw data remains faithful even as the Werner state approaches the pure Bell state $\ket{\Psi^-}$, i.e., as $\lambda \rightarrow 1$. For a pure Bell state, the MLE-computed bound underestimates the true bound by $4\%$.
\label{figureS2}}
\end{figure}
We have created 100 sets of simulated raw measurements for each $\lambda$ by assuming Gaussian amplifier noise consistent with the experiment. Figure \,S2 shows the concurrence bound $\mathcal{\B_{\psi-}}$ as a function of the true $\conc$ of the Werner state, obtained with and without MLE processing of the simulated noisy data. We find that while the mean of  $\mathcal{\B_{\psi-}}$ estimated directly from the raw data is unbiased, the mean of the concurrence bound obtained with MLE becomes increasingly  biased the more pure the Werner state, i.e., the closer $\lambda$ is to unity. MLE underestimates the bound by  $1\%$ at $\conc=0.85$, and by $4\%$ at $\conc=1$.

\section{VI. Possible Systematic Errors}

There are a variety of higher order and systematic effects that affect the accuracy of the entanglement 
metrology at the few-percent level. For example, the discrepancies between the experiment and master equation simulation in Fig.\,4E can arise from a systematic under-rotation of both qubits by only 1\%. There are also  higher order couplings that are relevant at this level. The first is the finite strength of the two-qubit $ZZ$ entangling interaction even in the off state ($\zeta/2\pi\sim1.2\,\MHz$ \cite{dicarlo_2009}). This residual coupling leads to errors in some of the two-qubit correlators on the order of $\zeta/\Omega^{\text{L(R)}}\sim2\%$. A second is the presence of the qubit-qubit swap interaction ($J/2\pi\sim 60~\MHz$ \cite{majer_coupling_2007}), leading to errors of order $J/(\wL-\wR)\sim 4\%$. Another effect is the qubit-state dependent filtering of the drive applied to a qubit, which is expected to be on the order of $\chi^{\text{R(L)}}/(\omega^{\text{L(R)}}-\wcav)\sim 2\%$. The effect of these couplings can be mitigated by implementing appropriate composite pulse schemes \cite{PhysRevA.67.042308} and will be explored in the future.